%% file: RCW86_v3.tex
\documentclass[useAMS,usenatbib]{mn2e}
\usepackage{journalabbr}
\usepackage{graphicx}
\usepackage{graphpap}

\newcommand{\halpha}{H$\alpha$}

\title[Optical proper motion measurements of RCW 86]{Proper Motions of H$\alpha$ filaments in the Supernova Remnant RCW 86 }
\author[E.A. Helder et al.]{E.A. Helder$^{1}$\thanks{E-mail:
helder@psu.edu}, J. Vink$^{2}$, A. Bamba$^{3}$, J.A.M. Bleeker$^{4}$, \newauthor
D.N. Burrows$^{1}$, P. Ghavamian$^{5}$ and R. Yamazaki$^{3}$ \\
$^{1}$the Pennsylvania State University, 525 Davey Lab, University Park, PA 16802, USA\\
$^{2}$Anton Pannekoek Institute/GRAPPA, University of Amsterdam, PO Box
94249, 1090 GE Amsterdam, The Netherlands\\
$^{3}$Department of Physics and Mathematics, College of Science and Engineering, Aoyama Gakuin University, 5-10-1 Fuchinobe, \\
Chuo-ku, Sagamihara, Kanagawa 252-5258, Japan\\
$^{4}$SRON Netherlands Institute for Space Research, Utrecht, The Netherlands\\
$^{5}$Department of Physics, Astronomy and Geosciences, Towson University, Towson, MD 21252, USA}
\begin{document}

\date{\today}

\pagerange{\pageref{firstpage}--\pageref{lastpage}} \pubyear{2012}

\maketitle

\label{firstpage}

\begin{abstract}

We present a proper motion study of the eastern shock-region of the
supernova remnant RCW 86 (MSH 14-6{\it 3}, G315.4-2.3), based on optical observations carried out
with VLT/FORS2 in 2007 and 2010.
For both the northeastern and southeastern regions, we measure an average
proper motion of H$\alpha$ filaments of 
$0.10 \pm 0.02$\arcsec\ yr$^{-1}$, corresponding to
$1200 \pm 200$~km\,s$^{-1}$ at 2.5~kpc.
There is substantial variation in the  derived proper motions, 
indicating shock velocities ranging from 
just below 700~km\,s$^{-1}$  to above 2200~km\,s$^{-1}$.

The optical proper motion is lower than the previously measured X-ray
proper motion of northeastern region. The new measurements
are consistent with the  previously measured proton
temperature of $2.3 \pm 0.3$~keV, assuming no  
cosmic-ray acceleration. However, within the uncertainties,
moderately efficient ($<$ 27 per cent) shock acceleration is still possible.
The combination of optical proper
motion and proton temperature rule out the possibility that RCW 86 
has a distance less than 1.5~kpc.

The similarity of the proper motions in the northeast and southeast
is peculiar, given the different densities and X-ray emission properties
of the regions. The northeastern region has lower densities and the X-ray emission is synchrotron dominated, suggesting that the shock velocities should
be higher than in the southeastern, thermal X-ray dominated, region.
A possible solution is that the
H$\alpha$ emitting filaments are biased toward denser regions, with lower
shock velocities. 
Alternatively,
in the northeast the shock velocity may have decreased rapidly
during the past 200~yr, 
and the X-ray synchrotron emission is an afterglow from a period when the
shock velocity was higher.

\end{abstract}

\begin{keywords}
interstellar matter -- optical: supernova remnants.
\end{keywords}

\section{Introduction}
For decades, supernova remnants have been considered the main accelerators of Galactic cosmic rays. One of the earliest arguments for this was their energy budget: in order to maintain the Galactic cosmic-ray density, supernovae need to put a substantial amount of their kinetic energy into accelerating cosmic rays. As there are not many sources capable of providing these amounts of energy, there are few other candidates \citep[for a list, see][]{Butt2009}. The idea that SNRs are the main sources of cosmic rays has been thoroughly investigated, both from the theoretical and observational perspective \citep[][for recent reviews]{Schure12,Helder12}. 

One method of testing whether supernova remnants are efficiently accelerating cosmic rays, is by investigating the energy budget from the hot plasma in the remnants. This method compares the energy in thermal particles right behind the shock front to the total kinetic energy available (deduced from the shock speed). If the available kinetic energy is not fully used for heating particles, one can attribute the energy deficit to energy in {\it non}-thermal particles \citep[e.g.,][]{Vink2010}. This method has been used to determine the cosmic-ray pressure behind the shock fronts of four remnants: 1E 0102-7219, the Cygnus Loop,  SNR 0509-67.5 and RCW 86,  \citep[][respectively]{Hughes0102,Salvesen2009,Helder2010,Helder2009}. For the first two remnants, the temperatures are based on the electron temperature measured from X-ray spectra. This method has the additional difficulty that the electron temperature does not necessarily provide an adequate measure of the mean post-shock plasma temperature \citep{Ghavamian2007}. 

\begin{figure}
\includegraphics[angle=0,width=0.45\textwidth]{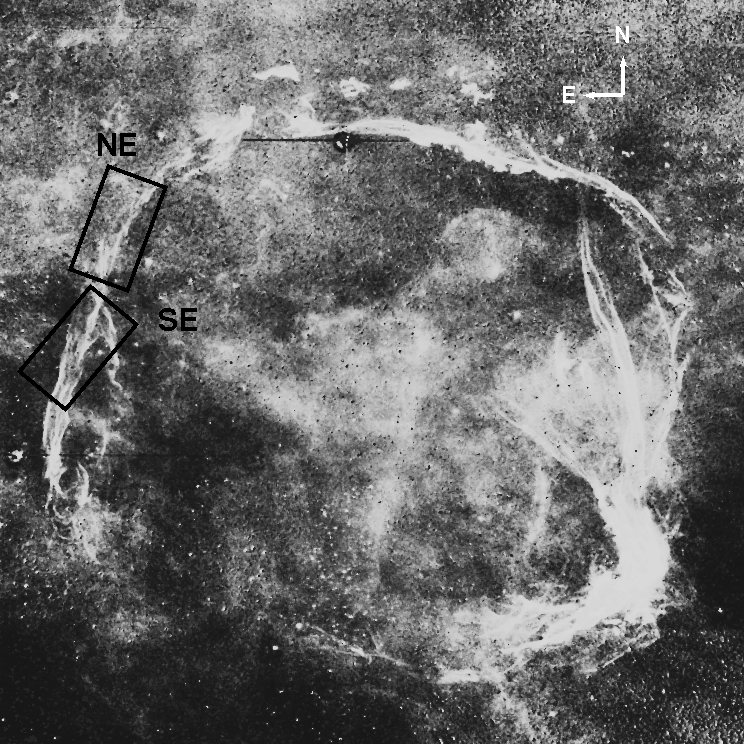}
\caption{H$\alpha$ image of the RCW 86 supernova remnant \citep{Smith1997}. Indicated are the locations of the pointings for which we determined proper motions in the northeast (NE) and southeast (SE), shown in Figure \ref{image}.}
\label{smith}
\end{figure}

The temperatures in the studies by \cite{Helder2010} and \cite{Helder2009} are based on the post-shock proton temperature. For Solar abundances, the proton temperature is more than half the mean plasma temperature. The proton temperatures in these studies were measured from the H$\alpha$ line emission. For non-radiative shocks in partially neutral gas, this H$\alpha$ line emission consists of both a narrow and a broad component superimposed. The narrow component is  caused by direct excitation of the neutral hydrogen after being swept up by the shock. 
The broad component is emitted after charge exchange between the swept-up neutral hydrogen atoms and the hot post-shock protons. The width of this component is therefore a measure of the post-shock proton temperature \citep[see][for a review]{Heng2010}.

A disadvantage of the shock velocity measurements for northeastern part of RCW 86 by 
 \citet[][]{Helder2009} are that they are based on X-ray proper motion measurements.
Because of the limited statistics of the
X-ray images the proper motions were measured for large regions. In contrast, the temperature measurements based on H$\alpha$ spectroscopy
are measured at very specific locations that are relatively bright in H$\alpha$. It is, therefore, preferable to combine H$\alpha$-based temperature
measurements with proper motions of  specific H$\alpha$ filaments.  For that reason, we obtained new \halpha\ images of the northeastern and southeastern parts of RCW~86 (MSH 14-6{\it 3}, G315.4-2.3, see Fig.~\ref{smith}) in order to measure the proper motions for those specific regions for which we had also measured the proton temperatures. We present those results here.

\begin{figure*}
\includegraphics[angle=0,width=0.8\textwidth]{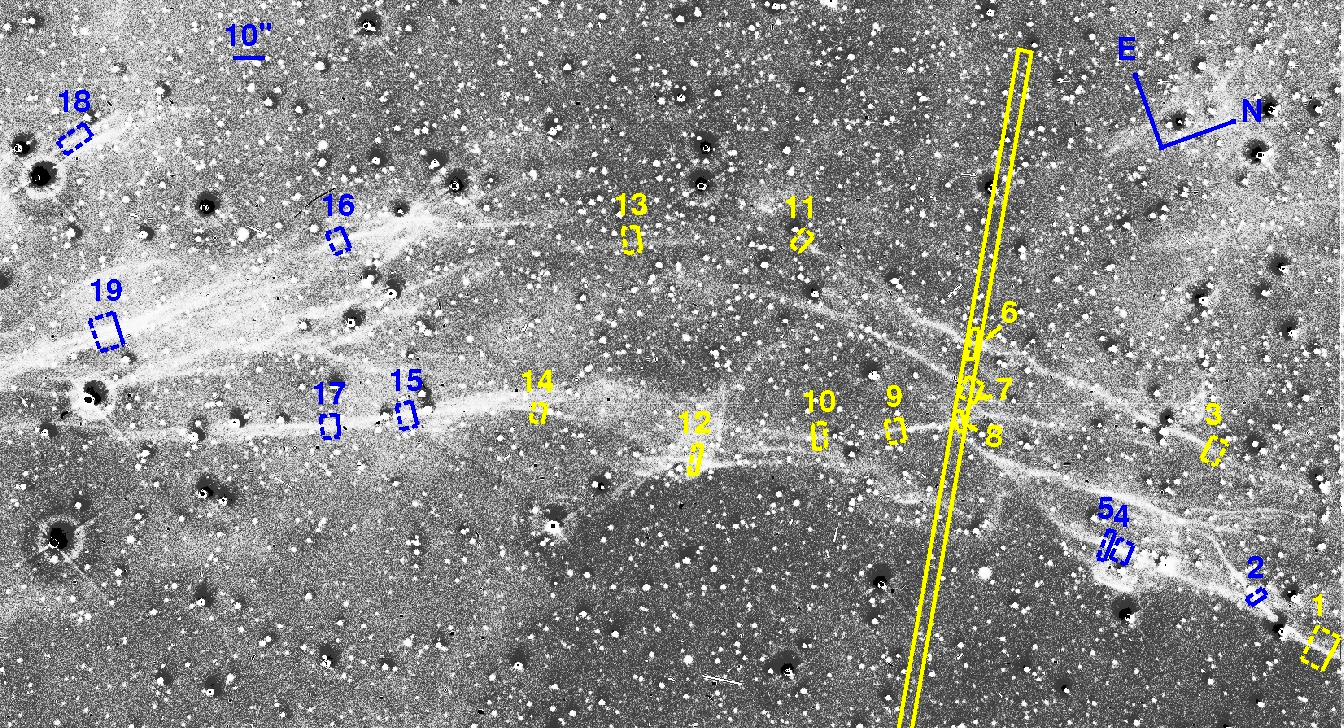}\\
\includegraphics[angle=0,width=0.8\textwidth]{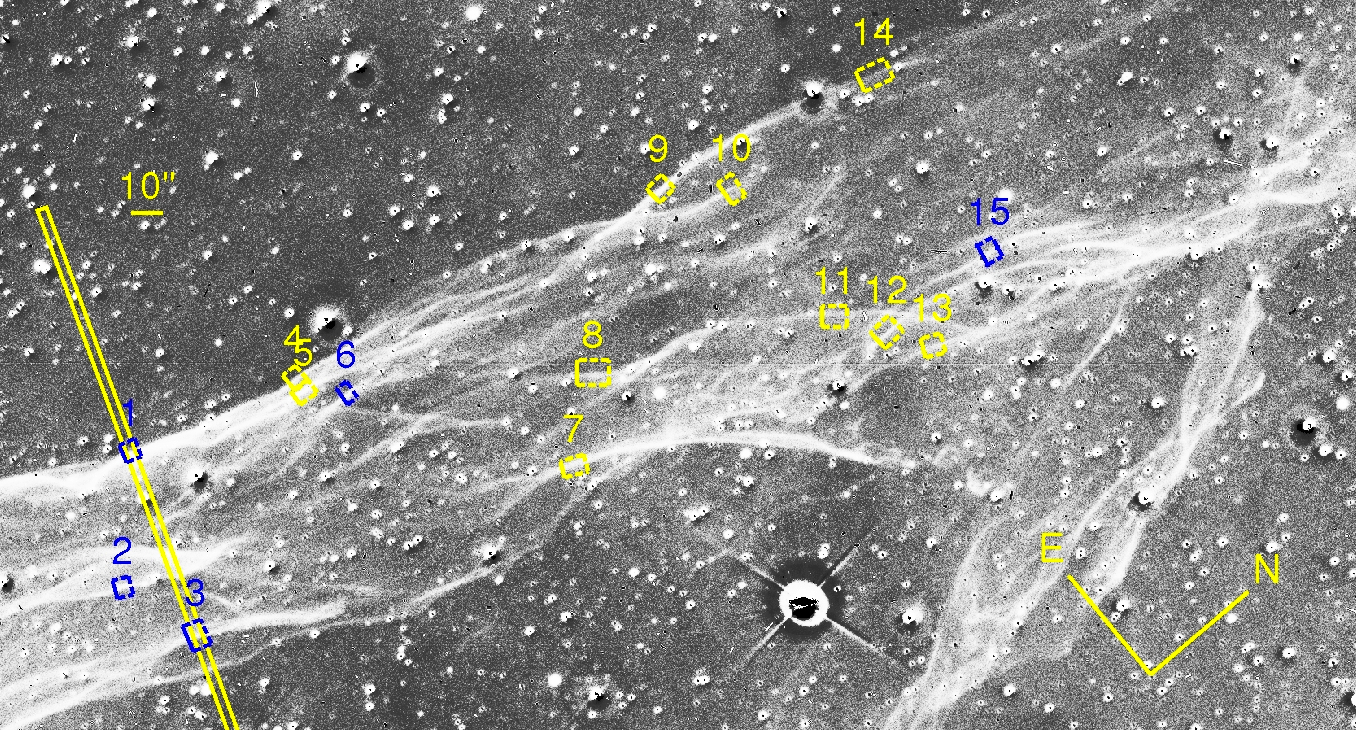}
\caption{VLT/FORS2 images of the northeast (top) and southeast (bottom) part of the remnant. Both observations were obtained in 2010 and both images have been continuum subtracted. North and east are indicated, as well as the scale and the regions we choose for determining proper motions (marked by the small boxes). The long rectangular boxes indicate the locations of the slits used in \citet{Helder2009,Helder2011}. For the top image, the centre of the remnant is below, for the bottom image, it is below and slightly to the right. 
The colours of the boxes were chosen to aid visibility in both lighter and darker parts of the image.}
\label{image}
\end{figure*}

\section[data]{Data and results}
\label{data}
The two epochs of data for this study were obtained with VLT/FORS2
\citep{FORS}. The first observation was done on 2007, February 25th
and the second observation on 2010, April 5th. This gives a time
baseline of 1135 days. For this study, we focus on the observations
centred on the northeast (RA: 14:45:11, Dec: -62:17:43, J2000) and
southeast (RA: 14:45:30, Dec: -62:25:21, J2000) of the remnant.  Both
observations include three images through an H$\alpha$ filter
(\texttt{H\_Alpha+83}) for 10 minutes in total per epoch as well as three images
through an H$\alpha$ filter with a velocity offset of 4500
km\,s$^{-1}$ (\texttt{H\_Alpha/4500+61}), also for a total of 10
minutes per pointing per epoch. The three frames were spatially dithered to account for
potential small-scale imperfections in the detector. 
The biases were subtracted and to correct
  for the uneven illumination across the detector we
used sky flats. Then the frames were combined for each epoch and filter. 
The individual images were aligned using the \texttt{interpol} routine of the image subtraction package ISIS \citep{Alard1998}. In this routine a two dimensional linear function is fitted to the positions of more than 350 stars per frame (exact number differs per frame) and the frames are remapped to match the corresponding stars in the reference frame. To subtract stars and other non-H$\alpha$ background emission, we subtracted images through the \texttt{H\_Alpha/4500+61} continuum filter from the images through the H$\alpha$ filter. The continuum images were scaled such that the resulting subtracted image would have the lowest standard deviation, bringing the subtracted sky emission down to zero. The resulting images are shown in Figure \ref{image}. We determined the scale of the images to be 0.252$''$/pixel by matching the images with the USNO-B1.0 catalog \citep{Monet03} from which we picked isolated stars. We utilised the same stars to estimate the spatial accuracy of our image matching. We determined the pixel coordinates of the centroids of these stars in both images. These coordinates all matched within 0.25 pixels for the northeast, and 0.12 pixels for the southeast. We conservatively consider these offsets to be the systematical error on our proper motion measurements.
\begin{figure*}
\includegraphics[angle=0,width=0.7\textwidth]{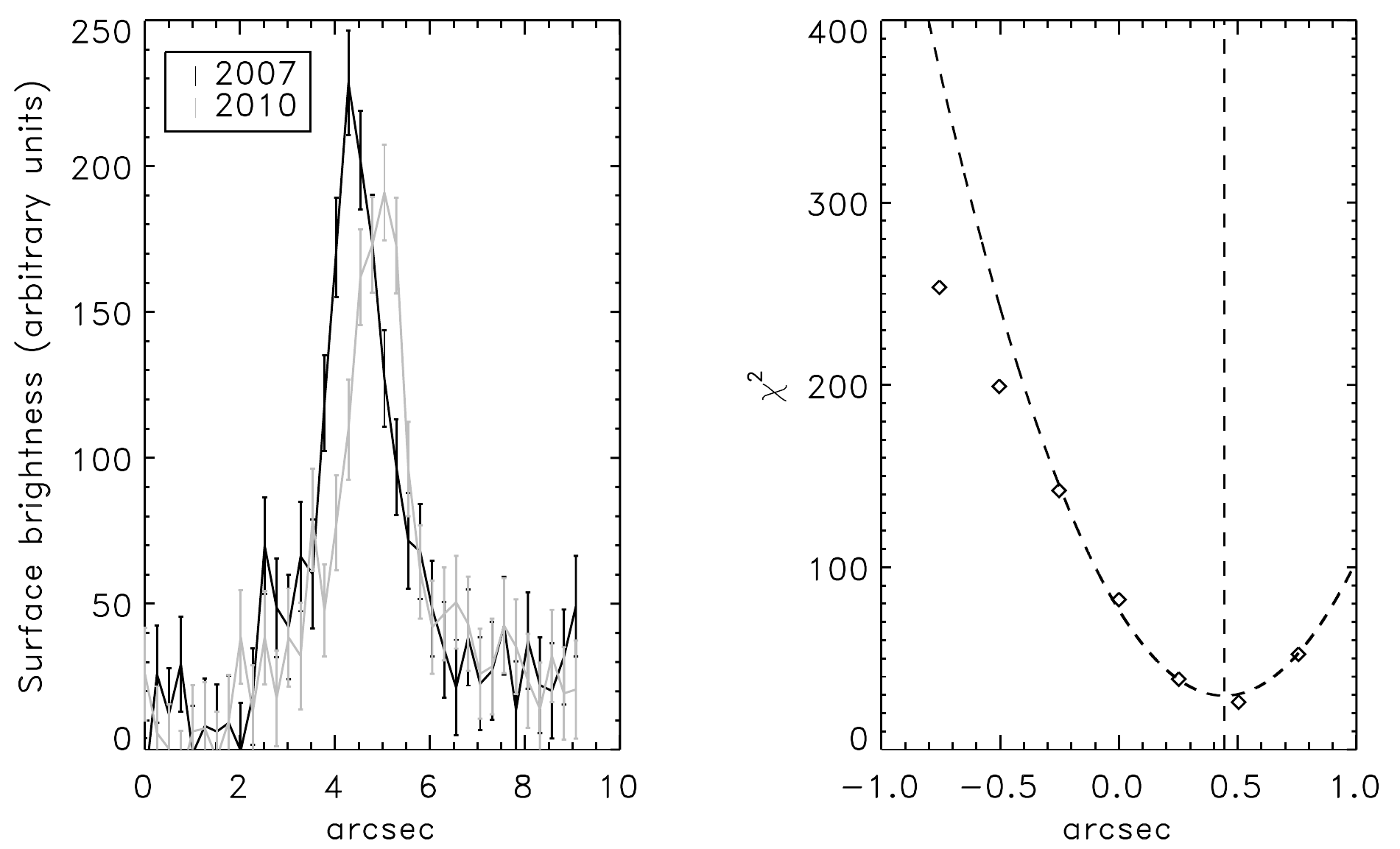}
\caption{{\em Left}: Surface brightness profiles taken in 2007 and 2010 of region 6 in the northeast part of the remnant. {\em Right}: $\chi^2$ values as function of filament displacement. The dashed line indicates the best-fit displacement. }
\label{shiftchi}
\end{figure*}
\input{shift_loop_NE_noxy.tex}
\input{shift_loop_SE_noxy.tex}
We made a mask for each image, to flag pixels that contain either cosmic-ray streaks or stars. We used a median filter for detecting cosmic rays and faint stars, and we set a maximum luminosity to detect bright stars.

Since RCW~86 is located in the Galactic plane, the field is crowded with background stars. To measure the proper motions as accurately as possible, we avoided filaments with stars in their close vicinity. Keeping this in mind, we selected regions across several filaments. We also made sure to cover the filaments that were used for the spectra described in \cite{Helder2009} and \citet[][regions 1 and 3 for the southeast and regions 6, 7 and 8 for the northeast, Figure \ref{image}]{Helder2011}. We extracted surface brightness profiles from these filaments. To correct for intrinsic brightness variations of the background of the images, we determined and subtracted the background in both the 2007 and 2010 profiles independently. 

To calculate the proper motions, we shifted the profiles over one another in steps of 1 pixel, calculating the $\chi^2$ for each shift, assuming a constant error for all bins (Figure \ref{shiftchi}). We determine the best-fit shift by fitting a parabola to the 7 $\chi^2$ values surrounding the minimal $\chi^2$. The surface brightness uncertainties were estimated, iteratively, from the dispersion of the residuals around the best fit model. This results per definition in $\chi^2_{\rm red} = 1$. 
Utilising these surface-brightness uncertainty estimates, we estimate the 1-$\sigma$ uncertainties on the best-fit proper motion, which
correspond to $\Delta \chi^2=1$. 
Tables \ref{table_SE} and \ref{table_NE} list the proper motions and 1-$\sigma$ uncertainties resulting from this procedure.

\begin{figure}
\includegraphics[angle=0,width=0.45\textwidth]{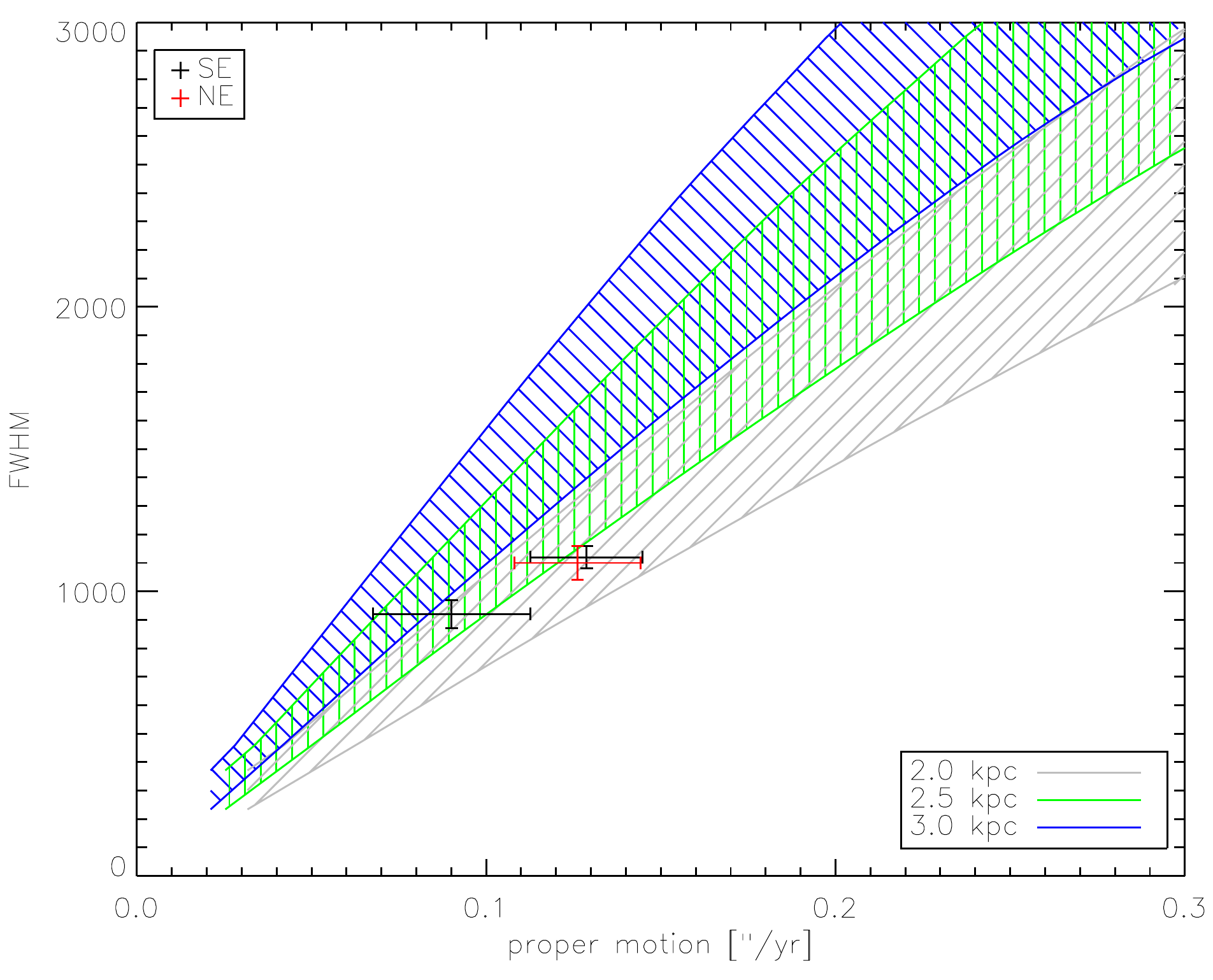}
\caption{
Proper motion measurements of the southeast (black) and northeast (red) filaments combined with the measurements
of \citet{Helder2009} of the FWHM of the broad components for these filaments. 
Overplotted are the line widths (FWHM) as function of velocity for a range of distances \citep{Adelsberg}. 
\label{distance}
}
\end{figure}

\def\kms{km\,s$^{-1}$}

\section{Discussion}
\label{discussion}

The \halpha\ proper motion measurements for the eastern region of RCW 86 
reported in our study provide some interesting results. First, the proper motions in the northeast 
do not agree with the X-ray proper motion of the X-ray synchrotron region.
The X-ray proper motion reported by \citet{Helder2009} is
$0.5\pm 0.2$\arcsec\ yr$^{-1}$, whereas for the northeast the error-weighted average
of all measured optical proper motions
is $0.095\pm0.03$\arcsec\ yr$^{-1}$. This is a 2-$\sigma$ deviation from the
X-ray proper motion. 

Second, we do not find a substantial difference in optical proper motion between
the northeastern, X-ray synchrotron-emitting, part of the remnant shell and the southeastern
shell, whose X-ray emission only shows evidence for thermal X-ray emission
\citep[e.g.][]{Vink2006}. For the southeast the error weighted average proper motion
is $0.10\pm 0.02$\arcsec\ yr$^{-1}$. 

\subsection{Implications for shock acceleration efficiency}
Both findings have implications for the particle acceleration properties
of RCW 86.
For a nominal distance of RCW 86 of 2.5~kpc, a proper motion of 
0.10\arcsec\ yr$^{-1}$ corresponds to a shock velocity of 1180~\kms.
A shock with this velocity is expected to heat the protons
to at least $kT_{\rm expected}=(3/16) \mu m_{\rm p}V_{\rm s}^2= 1.6 $~keV, with $\mu=0.6$ in case of full electron-ion
equilibration and $\mu=1$ if protons and electrons are completely unequilibrated (i.e. the temperature is proportional
to the particle mass, and all species have equal velocity distribution instead of equal energy distributions).
This is consistent with the proton temperature of $kT_{\rm p}=2.3\pm 0.3$~keV 
based on the width of the broad \halpha\ line reported by \citet{Helder2009},
as illustrated in Fig.~\ref{distance}.
Note that  for the southeastern part \citet{Helder2011} published two temperature measurements. For the northeastern part only one temperature was
published.

{\it Since there is no longer a discrepancy between the temperature expected from
the proper motion and  the measured proton temperature, there is also no
need to infer that a significant fraction of the shock's energy flux has been lost to cosmic-ray acceleration. }

However there is sufficient uncertainty
in the distance and proper motion to allow still for the possibility of
(moderately)
efficient shock acceleration: at the slit position for which the
proton temperature was measured  (filaments 6, 7, 8) the proper motion
is $0.13\pm0.02$\arcsec\ yr$^{-1}$ (error weighted average, including systematic errors).
If the distance to RCW 86 is at the high-end of the estimates, 3~kpc,
this corresponds to $1792\pm258$~\kms. This translates into an upper limit
on the ratio between measured and expected post-shock temperature of
$\beta\equiv kT_{\rm p}/kT_{\rm expected}>0.61$,
corresponding
to a limit on the post-shock cosmic-ray pressure of 
$w\equiv P_{\rm cr}/P_{\rm tot}<27$ per cent, 
according to the two-fluid model of \citet{Vink2010}. 
The 2-$\sigma$ error on the reported proper motion  allows in principle
even for an upper limit on the shock velocity of $<2300$~\kms, corresponding
to $kT_{\rm p} < 6.2$~keV, $\beta>0.37$ and an efficiency of
$w<46$ per cent.

On the other hand the combination of measured proton temperature
\citep{Helder2009} and the proper motions reported here 
reinforces the distance estimates of 
$2.5\pm 0.5$~kpc \citep{Rosado,Sollerman2003}, and make a distance
between 1.0-1.5~kpc \citep{Long,Bocchino2000} less likely, 
since for the measured proton temperature of $2.3\pm 0.3$ keV, a shock
velocity is required of at least 1080$\pm$70 \kms. 
This is at odds with our proper motion measurements, which for 1.0-1.5 kpc
imply shock velocities of  only 400-700 \kms.

\subsection{Implications for the X-ray synchrotron emission}
Although the efficient cosmic-ray acceleration in the northeast
of RCW 86 is no longer necessary to explain the observations, the \halpha\ proper motions pose some new
challenges for understanding the properties of this supernova remnant.

First, the X-ray emission properties of the northeastern
and southeastern filaments are quite distinct. The northeast is dominated by X-ray synchrotron
emission with only weak X-ray line emission from part of the region.
The weak line emission indicates a low ionisation age, implying a combination of
low density and/or recently heated plasma. 
In contrast, the southeast
filaments do not show any signs of X-ray synchrotron radiation, but have much
brighter thermal X-ray emission than the northeast, indicating higher plasma
densities and also larger ionisation ages \citep{Vink2006}. The \halpha\ emission in the northeast is also
fainter than in the southeast, consistent with the idea that in the northeast the shock
moves through a lower density medium than in the southeast. Given the apparent
contrast in densities, one expects also distinct shock velocities,
as the shock is expected to slow down in denser regions. This is inconsistent with the low and similar H$\alpha$ proper motions measured for the northeast and southeast.

The presence of X-ray synchrotron emission in the northeast is consistent
with relatively high shock velocities. This was
the reason for investigating the shock acceleration properties of the northeast
shock by \citet{Helder2009}.
Diffusive shock acceleration models
show that for synchrotron radiation with maximum photon energy limited by
radiative losses, the cut-off photon energy is independent of magnetic
field and scales as \citep{Zirakashvili,Vink2012}
\begin{equation}
h\nu_{\rm cut-off} \approx 1.4 \eta^{-1} 
\Bigl(\frac{V_{\rm s}}{5000\ {\rm km\,s^{-1}}}\Bigr)^2\ {\rm keV},
\label{eq:syn_max}
\end{equation}
with $\eta$ ($\geq$ 1) an efficiency factor with respect to optimal Bohm-diffusion
\citep[e.g.][]{Reynolds}. Indeed most supernova remnant shocks for which X-ray synchrotron
radiation has been detected appear to have shock velocities in excess
of 3000~\kms\ \citep[][]{Helder12}. 
Clearly the \halpha\ proper motions reported here do not agree
with this general trend, as for a distance of
2.5~kpc the proper motions should correspond to
to $h\nu_{\rm cut-off}<0.1$~keV. This low value is consistent with the
lack of X-ray synchrotron radiation in the southeast, but is at odds
with the prominence of X-ray synchrotron radiation from the northeast.
Note that also the H.E.S.S. TeV $\gamma$-ray map of RCW 86 indicates a lack
of TeV emission from the southeast \citep{AharonianRCW}.

Here, we offer two possible explanations for the presence of the
X-ray synchrotron radiation from the northeastern region and its
absence from the southeastern region.

First, it is possible that the overall shock velocity in the northeast
is not well represented by the \halpha\ proper motions, as suggested by \cite{Williams11}. Note that it is not uncommon for supernova shock speed measurements at different wave lengths to give contradictory answers in different wavelengths \citep{Moffett}. RCW 86 is characterised
by large density contrasts, so perhaps we only detect \halpha\ emission where
locally the shock encounters denser gas, and (temporarily) slows down, thereby
increasing in \halpha\ brightness. 
In this context it is interesting to note that all along the southeastern shock \halpha\
emission is present, whereas the northeastern shock region does not
show evidence for \halpha\ everywhere along the shock region.
The strong variation in densities could in principle mean that the overall
shock velocity in the northeast is closer to 3000~\kms, consistent with
the X-ray proper motions reported by \citet{Helder2009}, but that the
\halpha\ emitting regions are much slower. Indeed, the 
measured proper motions show  variation that cannot be explained
by statistical measurement errors. For example, filament 18 has a proper motion
that is 2-$\sigma$ higher than the average proper motion, and indicates that
locally the shock velocity may be higher than 2000~\kms. In contrast,
filament 1 corresponds to a 2.5-$\sigma$ deviation on the low 
side $V_{\rm s}\approx 750~$\kms.
The problem of explaining the X-ray/optical proper motion discrepancy in the northeast with large variations in shock velocities is that shocks in the southeast also show significant variation in proper motion. In order to test this explanation it
is important to obtain another X-ray proper motion measurement, now with
a longer baseline and therefore reduced measurement error. 

The second, alternative, explanation is that the \halpha\ proper motions
reported here are representative of the overall shock velocity in both the northeastern and southeastern regions, but that in the northeast the shock velocity
was much higher in the recent past. \citet{Vink2006} reported for this region
a magnetic field of $B\approx 26~\mu$G, consistent with the interpretation of GeV and 
TeV $\gamma$-ray observations of RCW 86 \citep{AharonianRCW,Lemoine12}. For these  magnetic fields, a relativistic electron with an energy of
$\sim$100 TeV has a synchrotron loss time of $\sim$180 yr. Therefore, it is possible
that 200~yr ago, the shock velocity was higher than 3000~\kms\
and nowadays, the region is still glowing in X-ray synchrotron radiation. 
This does not necessarily imply that there was a problem with the X-ray proper motion, as the X-ray proper motion
measures the velocity of the downstream plasma of the whole X-ray shell in the northeast, rather than the shock velocity.
Although a long X-ray synchrotron loss time in combination with a low shock velocity 
offers an explanation for the X-ray synchrotron emission from
the northeast, it does not offer an explanation for the absence of
X-ray synchrotron radiation from the southeast. Given the higher density in the southeast
the deceleration of the shock in the southeast should have been more severe,
whereas the \halpha\ proper motions for the two regions are comparable.

To complicate matters, one could also invoke magnetic field orientation as an additional ingredient for the presence or absence of X-ray synchrotron radiation. For example the geometry of the X-ray synchrotron emission from SN~1006 suggests that in this supernova remnant X-ray synchrotron emission only occurs when the magnetic field is parallel to the shock normal \citep{Rothenflug2004}. In RCW 86 the X-ray synchrotron emitting regions are roughly southwest-northeast aligned, more or less parallel to the Galactic plane. However, the morphology is not as clear as for SN~1006, and it does not agree with the idea that in the southwest the X-ray synchrotron emission is associated with the reverse shock, rather than with the forward shock \citep{Rho2002}.

A possible explanation for the different X-ray properties of the southeastern and northeastern regions could be offered by the hydrodynamic
simulations of supernova remnants evolving in a wind blown cavity \citep{Dwarkadas2005}. 
RCW 86 is regarded to be such a supernova remnant \citep{Vink1997,Vink2006,Williams11}. The simulations indicate that the shock velocity rapidly decreases once it starts interacting with the shell surrounding the cavity, but it recovers some of the shock velocity once it has penetrated the shell. The reason is that the initial interaction of the shock wave with the shell results in a slowing down of the shock, but as more of the material behind the shock transfers energy and momentum to the shell, it speeds up again. In the simulation presented
by \citet[][case 2]{Dwarkadas2005} the shell is encountered around 4000~yr, and as 
a result the shock velocity drops from $\sim 3000$~\kms, to a few hundred \kms,
but within a few hundred years the shock regains speed and 
continues with $\sim 1000$~\kms. 
These numbers are for the specific case
simulated, but it is conceivable that in different regions of RCW 86 the supernova remnant
shock is in different stages of interacting with the shell. In the northeast the shock could be approaching or just hitting the shell, and could therefore be rapidly
decelerating. In the southeast this could have happened in the more distant past, and the shock may actually have picked-up speed again. In such a situation one can still expect
X-ray synchrotron radiation in the northeast, but in southeast the shock slowed down too long ago
to still have electrons present with energies in excess of 10~TeV.

This interpretation is not entirely satisfying, as one would expect somewhere
on the eastern side to find a region that is in the stage of the interaction with the 
shell with the slowest shock velocities. Inspecting the variation in velocity
over the filaments there is not an obvious south-north trend. 
Instead, we find that in the northeast the inner filaments 
(northeast 1, 2, 4, 10, 12, 14) to have the lowest velocities.

\section{Conclusions}
\label{conclusions}
We measured proper motions of the H$\alpha$ emitting shock fronts on the east side of the RCW~86 supernova remnant. Based on our study, we reach the following conclusions: 
\begin{itemize}
\item[-] The shock velocities in the east part of the remnant display a large spread, varying from $>$ 700 km\,s$^{-1}$ to above 2200 km\,s$^{-1}$ assuming a distance of 2.5 kpc.
\item[-] These shock velocities are consistent with what we would expect from the proton temperatures determined from the broad H$\alpha$ lines in the southeast and northeast by \cite{Helder2009,Helder2011,Ghavamian2001} and \cite{Ghavamian2007}.
\item[-] Assuming a distance of 2.5 kpc, we do not need any cosmic-ray pressure to explain the proton temperature in the northeast of RCW 86.
\item[-] Taking into account measurement and distance uncertainties, a moderately efficient shock acceleration is still possible, with $<$ 27 per cent of the post-shock pressure being provided by accelerated particles. 
\item[-] The combination of the proper motion and the post-shock proton temperature rules out a distance of less than 1.5 kpc for this remnant.
\item[-] The proper motions of the filaments in the southeast and northeast are similar, which is surprising, given the very different nature of the dominant X-ray emission mechanism of these regions, namely synchrotron in the northeast and thermal X-ray emission in the southeast. We discuss two possible explanations for this. The first is that the \halpha\ emitting parts of the shocks have slowed down, whereas the X-ray synchrotron emitting shocks are still moving fast through the ambient medium, as suggested by \cite{Williams11}. The second explanation is that the X-ray expansion measurement was inaccurate, and the entire northeast shock has slowed down recently by interaction with a dense shell, and is still glowing in X-ray synchrotron radiation. The southeast shock has also slowed down by the same shell but has already overcome this shell and the shock velocity has increased again.
The two scenarios can be distinguished by obtaining a more accurate X-ray proper motion measurement with a new  Chandra observation, providing a much longer baseline.
\end{itemize}

\section*{Acknowledgments}

E.A.H. expresses her gratitude to Ana Chies Santos and Angela Adamo for discussions on the reduction of optical data. E.A.H. and D.N.B. are supported by SAO grants GO1-12070X and GO2-13064X. This research is based on observations collected with ESO telescopes at the Paranal Observatory under progamme IDs 079.D-0735(A) and 385.D-0483(A).


\label{lastpage}

\end{document}

%% file: shift_loop_NE_noxy.tex
\begin{center}
\begin{table}
\begin{tabular}{llrcl}
Filament \#  & shift [\arcsec] &  V$_{\rm s}$ &@& 2.5\,kpc\\
 &  in 1135 days & km\,s$^{-1}$ & $\pm$& stat. err. \\
\hline \\
 1 & 0.20 $\pm$ 0.04& 745 & $\pm$ & 136\\
 2 & 0.14 $\pm$ 0.07& 543 & $\pm$ & 280\\
 3 & 0.31 $\pm$ 0.09& 1172 & $\pm$ & 347\\
 4 & 0.25 $\pm$ 0.05& 948 & $\pm$ & 174\\
 5 & 0.28 $\pm$ 0.05& 1067 & $\pm$ & 186\\
 6 & 0.49 $\pm$ 0.07& 1871 & $\pm$ & 250\\
 7 & 0.31 $\pm$ 0.10& 1196 & $\pm$ & 367\\
 8 & 0.35 $\pm$ 0.06& 1325 & $\pm$ & 221\\
 9 & 0.34 $\pm$ 0.05& 1299 & $\pm$ & 191\\
 10 & 0.08 $\pm$ 0.11& 317 & $\pm$ & 437\\
 11 & 0.31 $\pm$ 0.09& 1192 & $\pm$ & 351\\
 12 & 0.26 $\pm$ 0.04& 991 & $\pm$ & 133\\
 13 & 0.39 $\pm$ 0.12& 1493 & $\pm$ & 475\\
 14 & 0.21 $\pm$ 0.10& 800 & $\pm$ & 371\\
 15 & 0.26 $\pm$ 0.07& 1001 & $\pm$ & 256\\
 16 & 0.37 $\pm$ 0.05& 1422 & $\pm$ & 175\\
 17& 0.29 $\pm$ 0.06& 1096 & $\pm$ & 219\\
 18 & 0.81 $\pm$ 0.23& 3071 & $\pm$ & 878\\
 19& 0.35 $\pm$ 0.04& 1349 & $\pm$ & 151\\
\noalign{\smallskip}
Mean/std. dev. &    0.31/0.08       &  1204&/&575 \\
\end{tabular}\\
\vskip 0.2mm
\caption{Proper motions for the northeast pointings. Quoted uncertainties are the statistical 1-$\sigma$ uncertainties. Systematic uncertainties are 0.063\arcsec and 240 km\,s$^{-1}$. 
The last row contains the average values and standard deviation. \label{table_NE}.}
\end{table}
\end{center}

%% file: shift_loop_SE_noxy.tex
\begin{center}
\begin{table}
\begin{tabular}{llrcl}
Filament \#  & shift [\arcsec] & V$_{\rm s}$ &@& 2.5\,kpc\\
 &  in 1135 days & km\,s$^{-1}$ & $\pm$& stat. err.  \\
 \hline \\
 1& 0.40 $\pm$ 0.04& 1531 & $\pm$ & 144\\
 2 & 0.13 $\pm$ 0.04& 478 & $\pm$ & 137\\
 3 & 0.28 $\pm$ 0.06& 1082 & $\pm$ & 225\\
 4 & 0.38 $\pm$ 0.05& 1446 & $\pm$ & 179\\
 5 & 0.22 $\pm$ 0.03& 852 & $\pm$ & 121\\
 6 & 0.35 $\pm$ 0.05& 1330 & $\pm$ & 175\\
 7 & 0.43 $\pm$ 0.06& 1653 & $\pm$ & 228\\
 8 & 0.38 $\pm$ 0.04& 1431 & $\pm$ & 137\\
 9 & 0.38 $\pm$ 0.04& 1444 & $\pm$ & 158\\
 10 & 0.37 $\pm$ 0.07& 1403 & $\pm$ & 256\\
 11 & 0.24 $\pm$ 0.13& 910 & $\pm$ & 501\\
 12 & 0.18 $\pm$ 0.12& 671 & $\pm$ & 448\\
 13 & 0.33 $\pm$ 0.06& 1246 & $\pm$ & 230\\
 14 & 0.49 $\pm$ 0.07& 1881 & $\pm$ & 248\\
 15 & 0.32 $\pm$ 0.08& 1213 & $\pm$ & 289\\
\noalign{\smallskip}
Mean/std. dev & 0.33/0.10 & 1240 & / & 374\\
\end{tabular}\\
\vskip 0.2mm
\caption{Proper motions for the southeast pointings. Quoted uncertainties are the statistical 1-$\sigma$ uncertainties. Systematic uncertainties are 0.03\arcsec and 114 km\,s$^{-1}$ \label{table_SE}}
\end{table}
\end{center}

%% file: RCW86_v3.bbl
\begin{thebibliography}{}

\bibitem[\protect\citeauthoryear{{Aharonian} et~al.,}{{Aharonian}
  et~al.}{2009}]{AharonianRCW}
{Aharonian} F.,  et~al., 2009, \apj, 692, 1500

\bibitem[\protect\citeauthoryear{{Alard} \& {Lupton}}{{Alard} \&
  {Lupton}}{1998}]{Alard1998}
{Alard} C.,  {Lupton} R.~H.,  1998, \apj, 503, 325

\bibitem[\protect\citeauthoryear{{Appenzeller} et~al.,}{{Appenzeller}
  et~al.}{1998}]{FORS}
{Appenzeller} I.,  et~al., 1998, The Messenger, 94, 1

\bibitem[\protect\citeauthoryear{{Bocchino}, {Vink}, {Favata}, {Maggio} \&
  {Sciortino}}{{Bocchino} et~al.}{2000}]{Bocchino2000}
{Bocchino} F.,  {Vink} J.,  {Favata} F.,  {Maggio} A.,    {Sciortino} S.,
  2000, \aap, 360, 671

\bibitem[\protect\citeauthoryear{{Butt}}{{Butt}}{2009}]{Butt2009}
{Butt} Y.,  2009, \nat, 460, 701

\bibitem[\protect\citeauthoryear{{Dwarkadas}}{{Dwarkadas}}{2005}]{Dwarkadas2005}
{Dwarkadas} V.~V.,  2005, \apj, 630, 892

\bibitem[\protect\citeauthoryear{{Ghavamian}, {Laming} \&
  {Rakowski}}{{Ghavamian} et~al.}{2007}]{Ghavamian2007}
{Ghavamian} P.,  {Laming} J.~M.,    {Rakowski} C.~E.,  2007, \apjl, 654, L69

\bibitem[\protect\citeauthoryear{{Ghavamian}, {Raymond}, {Smith} \&
  {Hartigan}}{{Ghavamian} et~al.}{2001}]{Ghavamian2001}
{Ghavamian} P.,  {Raymond} J.,  {Smith} R.~C.,    {Hartigan} P.,  2001, \apj,
  547, 995

\bibitem[\protect\citeauthoryear{{Helder}, {Kosenko} \& {Vink}}{{Helder}
  et~al.}{2010}]{Helder2010}
{Helder} E.~A.,  {Kosenko} D.,    {Vink} J.,  2010, \apjl, 719, L140

\bibitem[\protect\citeauthoryear{{Helder}, {Vink} \& {Bassa}}{{Helder}
  et~al.}{2011}]{Helder2011}
{Helder} E.~A.,  {Vink} J.,    {Bassa} C.~G.,  2011, \apj, 737, 85

\bibitem[\protect\citeauthoryear{{Helder}, {Vink}, {Bassa}, {Bamba}, {Bleeker},
  {Funk}, {Ghavamian}, {van der Heyden}, {Verbunt} \& {Yamazaki}}{{Helder}
  et~al.}{2009}]{Helder2009}
{Helder} E.~A.,  {Vink} J.,  {Bassa} C.~G.,  {Bamba} A.,  {Bleeker} J.~A.~M.,
  {Funk} S.,  {Ghavamian} P.,  {van der Heyden} K.~J.,  {Verbunt} F.,
  {Yamazaki} R.,  2009, Science, 325, 719

\bibitem[\protect\citeauthoryear{{Helder}, {Vink}, {Bykov}, {Ohira}, {Raymond}
  \& {Terrier}}{{Helder} et~al.}{2012}]{Helder12}
{Helder} E.~A.,  {Vink} J.,  {Bykov} A.~M.,  {Ohira} Y.,  {Raymond} J.~C.,
  {Terrier} R.,  2012, \ssr, 173, 369

\bibitem[\protect\citeauthoryear{{Heng}}{{Heng}}{2010}]{Heng2010}
{Heng} K.,  2010, PASA, 27, 23

\bibitem[\protect\citeauthoryear{{Hughes}, {Rakowski} \&
  {Decourchelle}}{{Hughes} et~al.}{2000}]{Hughes0102}
{Hughes} J.~P.,  {Rakowski} C.~E.,    {Decourchelle} A.,  2000, \apjl, 543, L61

\bibitem[\protect\citeauthoryear{{Lemoine-Goumard}, {Renaud}, {Vink}, {Allen},
  {Bamba}, {Giordano} \& {Uchiyama}}{{Lemoine-Goumard}
  et~al.}{2012}]{Lemoine12}
{Lemoine-Goumard} M.,  {Renaud} M.,  {Vink} J.,  {Allen} G.~E.,  {Bamba} A.,
  {Giordano} F.,    {Uchiyama} Y.,  2012, \aap, 545, A28

\bibitem[\protect\citeauthoryear{{Long} \& {Blair}}{{Long} \&
  {Blair}}{1990}]{Long}
{Long} K.~S.,  {Blair} W.~P.,  1990, \apjl, 358, L13

\bibitem[\protect\citeauthoryear{{Moffett}, {Goss} \& {Reynolds}}{{Moffett}
  et~al.}{1993}]{Moffett}
{Moffett} D.~A.,  {Goss} W.~M.,    {Reynolds} S.~P.,  1993, \aj, 106, 1566

\bibitem[\protect\citeauthoryear{{Monet} et~al.,}{{Monet}
  et~al.}{2003}]{Monet03}
{Monet} D.~G.,  et~al., 2003, \aj, 125, 984

\bibitem[\protect\citeauthoryear{{Reynolds}}{{Reynolds}}{1998}]{Reynolds}
{Reynolds} S.~P.,  1998, \apj, 493, 375

\bibitem[\protect\citeauthoryear{{Rho}, {Dyer}, {Borkowski} \&
  {Reynolds}}{{Rho} et~al.}{2002}]{Rho2002}
{Rho} J.,  {Dyer} K.~K.,  {Borkowski} K.~J.,    {Reynolds} S.~P.,  2002, \apj,
  581, 1116

\bibitem[\protect\citeauthoryear{{Rosado}, {Ambrocio-Cruz}, {Le Coarer} \&
  {Marcelin}}{{Rosado} et~al.}{1996}]{Rosado}
{Rosado} M.,  {Ambrocio-Cruz} P.,  {Le Coarer} E.,    {Marcelin} M.,  1996,
  \aap, 315, 243

\bibitem[\protect\citeauthoryear{{Rothenflug}, {Ballet}, {Dubner}, {Giacani},
  {Decourchelle} \& {Ferrando}}{{Rothenflug} et~al.}{2004}]{Rothenflug2004}
{Rothenflug} R.,  {Ballet} J.,  {Dubner} G.,  {Giacani} E.,  {Decourchelle} A.,
     {Ferrando} P.,  2004, \aap, 425, 121

\bibitem[\protect\citeauthoryear{{Salvesen}, {Raymond} \& {Edgar}}{{Salvesen}
  et~al.}{2009}]{Salvesen2009}
{Salvesen} G.,  {Raymond} J.~C.,    {Edgar} R.~J.,  2009, \apj, 702, 327

\bibitem[\protect\citeauthoryear{{Schure}, {Bell}, {O'C Drury} \&
  {Bykov}}{{Schure} et~al.}{2012}]{Schure12}
{Schure} K.~M.,  {Bell} A.~R.,  {O'C Drury} L.,    {Bykov} A.~M.,  2012, \ssr,
  173, 491

\bibitem[\protect\citeauthoryear{{Smith}}{{Smith}}{1997}]{Smith1997}
{Smith} R.~C.,  1997, \aj, 114, 2664

\bibitem[\protect\citeauthoryear{{Sollerman}, {Ghavamian}, {Lundqvist} \&
  {Smith}}{{Sollerman} et~al.}{2003}]{Sollerman2003}
{Sollerman} J.,  {Ghavamian} P.,  {Lundqvist} P.,    {Smith} R.~C.,  2003,
  \aap, 407, 249

\bibitem[\protect\citeauthoryear{{van Adelsberg}, {Heng}, {McCray} \&
  {Raymond}}{{van Adelsberg} et~al.}{2008}]{Adelsberg}
{van Adelsberg} M.,  {Heng} K.,  {McCray} R.,    {Raymond} J.~C.,  2008, \apj,
  689, 1089

\bibitem[\protect\citeauthoryear{{Vink}}{{Vink}}{2012}]{Vink2012}
{Vink} J.,  2012, \aapr, 20, 49

\bibitem[\protect\citeauthoryear{{Vink}, {Bleeker}, {van der Heyden}, {Bykov},
  {Bamba} \& {Yamazaki}}{{Vink} et~al.}{2006}]{Vink2006}
{Vink} J.,  {Bleeker} J.,  {van der Heyden} K.,  {Bykov} A.,  {Bamba} A.,
  {Yamazaki} R.,  2006, \apjl, 648, L33

\bibitem[\protect\citeauthoryear{{Vink}, {Kaastra} \& {Bleeker}}{{Vink}
  et~al.}{1997}]{Vink1997}
{Vink} J.,  {Kaastra} J.~S.,    {Bleeker} J.~A.~M.,  1997, \aap, 328, 628

\bibitem[\protect\citeauthoryear{{Vink}, {Yamazaki}, {Helder} \&
  {Schure}}{{Vink} et~al.}{2010}]{Vink2010}
{Vink} J.,  {Yamazaki} R.,  {Helder} E.~A.,    {Schure} K.~M.,  2010, \apj,
  722, 1727

\bibitem[\protect\citeauthoryear{{Williams}, {Blair}, {Blondin}, {Borkowski},
  {Ghavamian}, {Long}, {Raymond}, {Reynolds}, {Rho} \& {Winkler}}{{Williams}
  et~al.}{2011}]{Williams11}
{Williams} B.~J.,  {Blair} W.~P.,  {Blondin} J.~M.,  {Borkowski} K.~J.,
  {Ghavamian} P.,  {Long} K.~S.,  {Raymond} J.~C.,  {Reynolds} S.~P.,  {Rho}
  J.,    {Winkler} P.~F.,  2011, \apj, 741, 96

\bibitem[\protect\citeauthoryear{{Zirakashvili} \& {Aharonian}}{{Zirakashvili}
  \& {Aharonian}}{2007}]{Zirakashvili}
{Zirakashvili} V.~N.,  {Aharonian} F.,  2007, \aap, 465, 695

\end{thebibliography}
